\documentclass[3p,twocolumn]{elsarticle}
\pdfoutput=1
\usepackage{siunitx}
\usepackage{tikz}
\usepackage{amsmath}
\usepackage{subfig}
\usepackage{lineno}
\usepackage{booktabs}

\renewcommand{\vec}{\mathbf}
\newcommand*{\de}{\mathop{}\!\mathrm{d}}
\newcommand*{\uimm}{\mathrm{i}}

\begin{document}

\begin{frontmatter}

\title{Modelling of pulse train generation for resonant laser wakefield acceleration using a delay mask}

\author[ino]{G.~Vantaggiato}
\author[ino,infn]{L.~Labate\corref{ll}}\ead{luca.labate@ino.it}
\author[ino]{P.~Tomassini}
\author[ino,infn]{L.A.~Gizzi}

\address[ino]{Intense Laser Irradiation Laboratory, INO-CNR, Pisa, Italy}
\address[infn]{Istituto Nazionale di Fisica Nucleare, Sez. Pisa, Italy}

\cortext[ll]{Corresponding author}

\begin{abstract}
A new method for the generation of a train of pulses from a single high-energy, ultra short pulse is presented, suited for Resonant Multi-Pulse Ionization injection~\cite{art:toma17}. The method is based on different transverse portion of the pulse being delayed by a ``mask'' sectioned in concentric
zones with different thicknesses, in order to deliver multiple laser pulses. The mask is placed right before the last focusing parabola. A hole in
the middle of the mask lets part of the original pulse to pass through to drive
electron injection.
In this paper a full numerical modelling of this scheme is presented. In particular we discuss the spatial and temporal profile of the pulses emerging from the mask and how they are related to the radius and thickness of each section.
\end{abstract}


\end{frontmatter}

\section{Introduction}
\begin{figure}\centering
\includegraphics[width=.95\columnwidth]{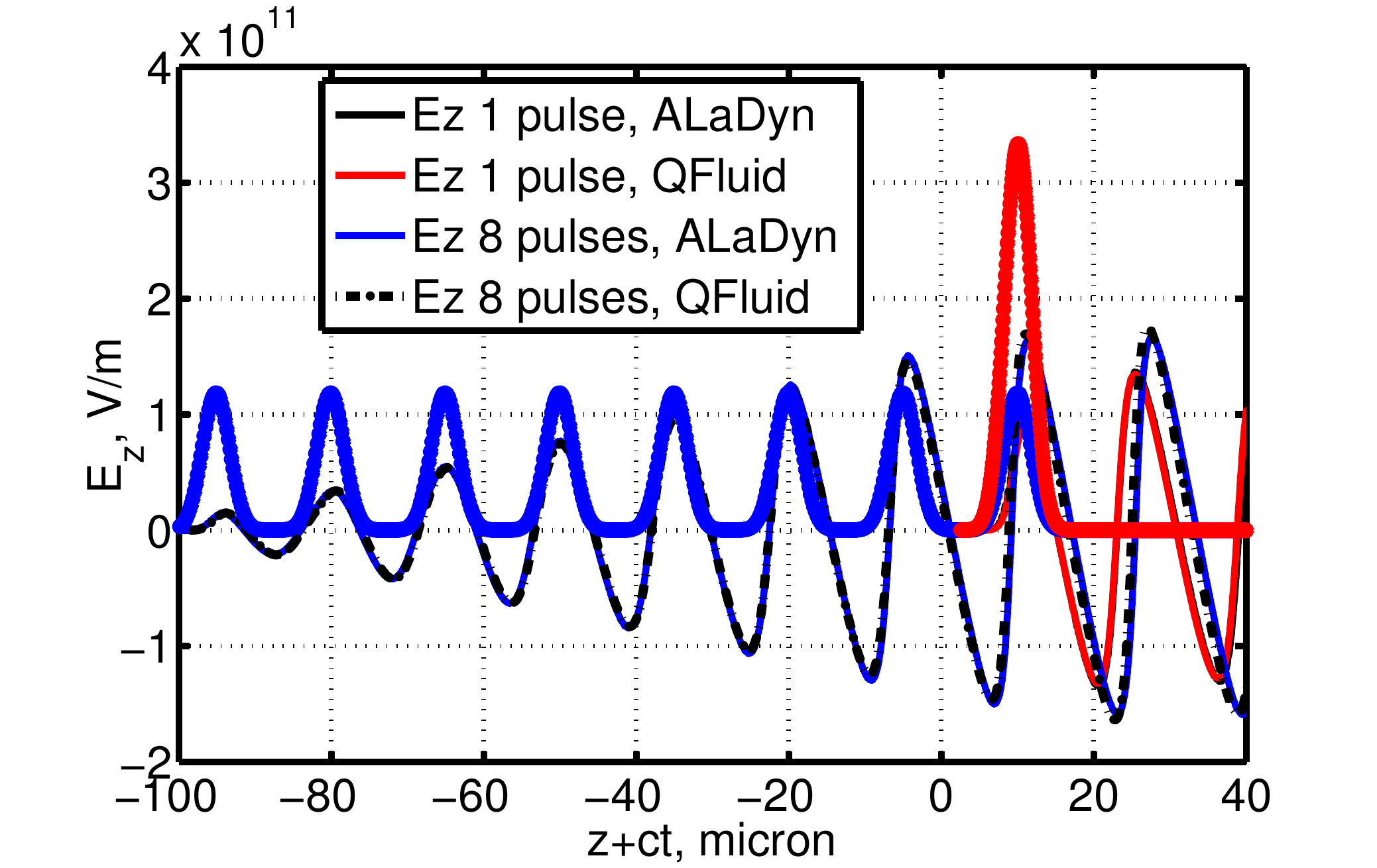}
\caption{Comparison of wakefield ectitation by a single pulse or a pulse train. In these plasma simulations pulses moving through the left are considered. The thin lines indicate the plasma wake. In this example a train of 8 pulses equally separated by the plasma period drives a wave (thin blue) whose maximum accelerating gradient is 20\% more than that of the wave excited by the single pulse (thin red) wiht the same delivered energy. Figure from \cite{art:toma17}.}\label{fig:mplwfa}
\end{figure}

Future applications of Laser WakeField Acceleration (LWFA), such as FELs and new generation of particle colliders, require high-quality electron bunches. The main approach consists in injecting low-emittance electron bunches in the plasma wave. Among the different models proposed, the Resonant  Multi-Pulse Ionization injection (ReMPI)~\cite{art:toma17} relies on the plasma wake being excited by a train of pulses produced from a single high-energy, ultra short laser pulse. The theory of plasma wave excitation by a pulse train was presented in \cite{art:umst94,art:umst95} (and more recently discussed in \cite{art:hook14}); the scheme takes advantage of the coherent sum of the wakefields produced by each pulse of the train (the pulses being separated in time by a plasma period) to excite a plasma wave with suitable amplitude. In Fig.~\ref{fig:mplwfa} a QFluid \cite{art:toma16} plasma simulation shows a comparison between this scheme and the classic LWFA. In this example a train of 8 pulses equally separated by a plasma period, each with duration of \SI{10}{fs} and peak intensity of \SI{7.4e17}{W/cm^2}, travels through a plasma with $n_e=\SI{5e18}{cm^{-3}}$ exciting a wave with a longitudinal accelerating field 20\% higher than the one generated by a single pulse moving through the same plasma with duration of \SI{10}{fs}, peak intensity of \SI{5.9e18}{W/cm^2} and the same delivered energy of the train. The advantage of the resonant scheme is that the reduced peak intensity of each train pulse enables controlled particle trapping conditions. Finally, a fraction of the original pulse is frequency doubled in order to provide separated ionization injection.

The generation of the pulse train from a single laser pulse has been widely investigated, using stacked Michelson interferometers \cite{art:side98,art:cowl17}, a linear array of birefringent plates \cite{art:drom07} and spectral filtering of a chirped pulse \cite{art:robi07,art:shal16}. All these techniques require challenging alignments and big optical elements due to the energy involved.

In this work we present a relatively easy-to-implement method of train generation, employing a structured transmission mask which allows different portions of the original beam to be delayed by different thicknesses of the transmissive layers. Such a simple configuration makes our technique suitable to produce pulse trains with existing laser systems and small laboratories. We developed a complete characterization of both spatial and temporal profile of the train pulses using vector diffraction theory and nonlinear time dependent analysis of pulse propagation, respectively.

\section{Description of the experimental method and space and time characterization of the pulse train}
\begin{figure}[t]\centering
\includegraphics[width=.95\columnwidth]{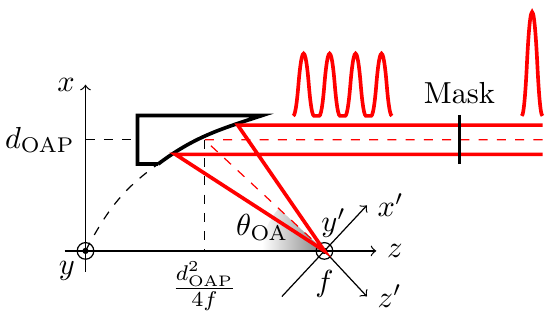}
\caption{Geometrical model. Cartesian coordinate system with the origin in the paraboloid vertex. The laser pulse comes from the right and splits in 4 rings separated in time that will be focused on the focal plane $x'y'$.}\label{fig:scheme}
\end{figure}

As shown in Fig.~\ref{fig:scheme} we consider a single high-energy pulse, with a super-gaussian transverse profile, impacting on a concentric-sectioned mask placed just before the last focusing Off-Axis Parabola (OAP).  Each section's thickness $d_i$ is chosen to match the time delay of the emerging ring pulses with the plasma period. Then these rings are focused on the plasma target and results in a multi-pulse driver for the wakefield.

\begin{figure}[t]\centering
\subfloat[3D-model.\label{fig:mask}]{\includegraphics[width=.3\columnwidth]{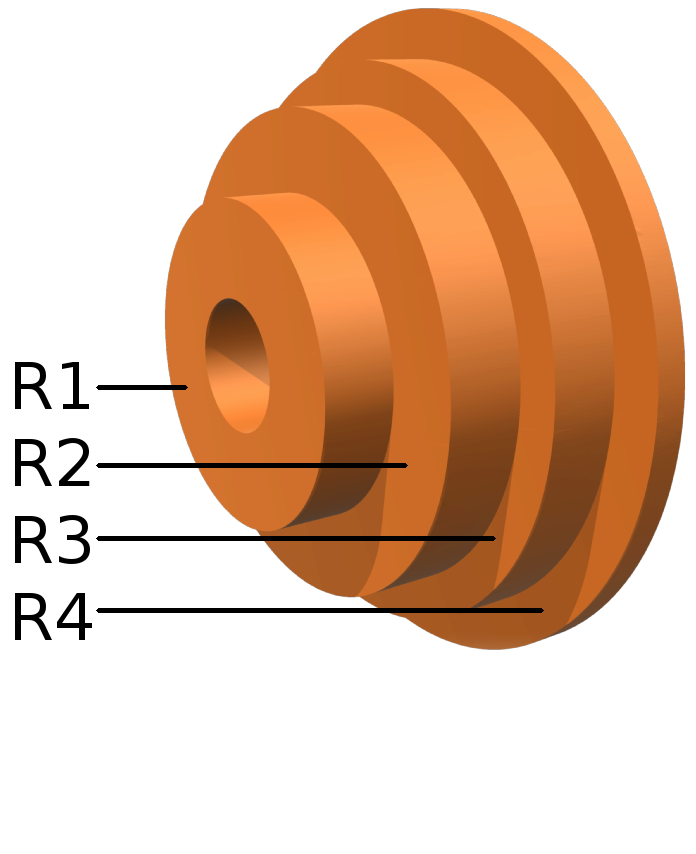}}
\subfloat[Transverse beam profile.\label{fig:radii}]{\includegraphics[width=.7\columnwidth]{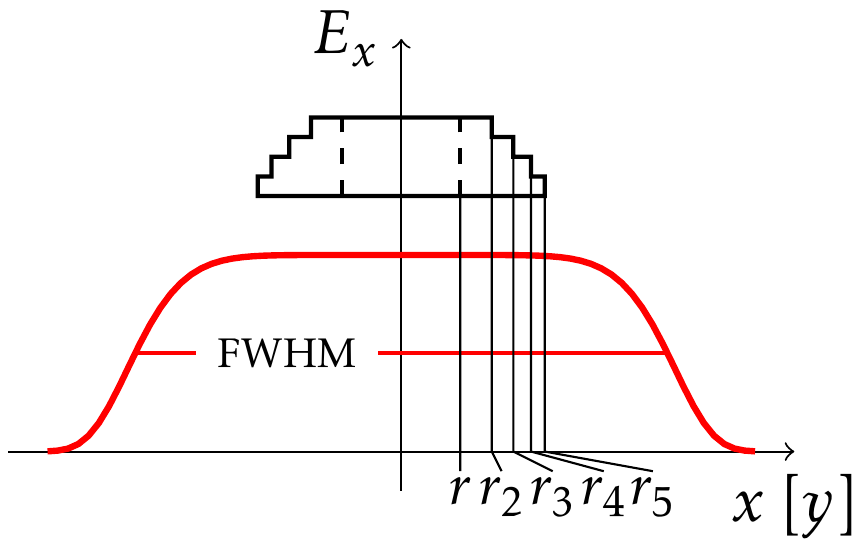}}
\caption{Mask and beam. The incoming beam with a super-gaussian transverse profile impacts normally on the mask and emerges as 4 rings time-separated. Sections with the same area give rise to pulses carrying the same energy. The hole in the middle is due to the necessity in ReMPI model to use part of the original pulse for ionization injection.}
\end{figure}

For a first case study we designed a delay mask suitable for the already mentioned ReMPI scheme. This mask (Fig.~\ref{fig:mask}) delivers a 4-pulses train while the hole in the middle allows the so-called ``ionization pulse'' to pass through. This pulse, with relatively low energy ($\sim \SI{100}{mJ}$), can be produced in many ways. For instance, a pick up mirror can be used (upstream of the delay mask) and a suitable delay line can be set up, together with a frequency doubling crystal; a second small mirror can be used to re-insert the pulse into the main path. Since this paper is mostly concerned with the generation of the pulse train, we defer to another paper a deeper discussion of this issue.

The internal and external radii of each section are initially set in order to carry the same energy, i.e. each region has the same area (considering that the whole mask lies in the flat region of the super-gaussian transverse profile, as in Fig.~\ref{fig:radii}). Therefore $i$-th radius is calculated as $r_i = \sqrt{i}r$ with $r$ being the central hole radius.

\subsection{Study of the spatial properties of the train pulses}
Each pulse of the train is spatially characterized by its electrical field at the focal plane $x'y'$ of the OAP. The calculations are performed in the framework of the Stratton-Chu vector diffraction theory. Time dependence is intentionally omitted in this analysis and contour effects are negligible in the far-field calculation.
We consider a super-gaussian transverse profile and linear polarization along the $x$ axis for the incoming beam that, with reference to the coordinate system depicted in Fig.~\ref{fig:scheme}, is written as
\begin{equation}
E(\vec{x})=\exp \bigg[-\frac{1}{2} \bigg(\Big(\frac{x-d_\text{OAP}}{\sigma_x}\Big)^2 + \Big(\frac{y}{\sigma_y}\Big)^2 \bigg)^4 \bigg]\,,
\end{equation}
with $\sigma_x\equiv\sigma_y=\textrm{FWHM}/2(\ln2)^{1/8}$, FWHM being the transverse size of the beam impinging on the mask, as represented in Fig.~\ref{fig:radii}. 
Following the same formulation of \citep{art:laba16} for the electric field at each point $\vec{x}_P$ of the focal plane we get
\begin{multline}\label{eqn:eint}
E_j(\vec{x}_P) = -\frac{1}{\lambda} \int_\text{OAP}  E(\vec{x}) \Bigl(g_j^{(r)}(\vec{x},\vec{x}_P) \sin \bigl(kv(\vec{x},\vec{x}_P)\bigr) \\
+  g_j^{(i)}(\vec{x},\vec{x}_P) \cos \bigl(kv(\vec{x},\vec{x}_P)\bigr)\Bigr)  \de x \de y\,,
\end{multline}
where $\lambda$ is the laser wavelength, $k$ the wave vector and $v = u(\vec{x},\vec{x}_P)+p(\vec{x})$ with $u=\lvert\vec{x}-\vec{x}_P\rvert$ being the distance from the surface point $\vec{x}$ to the point $\vec{x}_P$; $p=(d^2_\textrm{OAP} - x^2 - y^2)/4f$ is the difference of optical path with respect to the OAP center. $g_j^{(r)}$ and $g_j^{(i)}$ are the real and imaginary part, respectively, of the complex $g_j$ functions that can be written as:
\begin{subequations}
\begin{align}
g_x &= \frac{1}{u} - \frac{x}{2f} \biggl(1 - \frac{1}{\uimm k u}\biggr) \frac{x-x_P}{u^2}\\
g_y &= - \frac{x}{2f} \biggl(1 - \frac{1}{\uimm k u}\biggr) \frac{y-y_P}{u^2}\\
g_z &= \frac{1}{u}\frac{x}{2f} - \frac{x}{2f} \biggl(1 - \frac{1}{\uimm k u}\biggr) \frac{x-x_P}{u^2}\,.
\end{align}
\end{subequations}

In order to finally calculate the electric field we perform the numerical integration in equation~\eqref{eqn:eint} using a C++ code we developed. In particular, integrating on the portion of the OAP selected by a certain ring we get the intensity map of the corresponding train pulse.

\begin{figure}\centering
\includegraphics[width=\columnwidth]{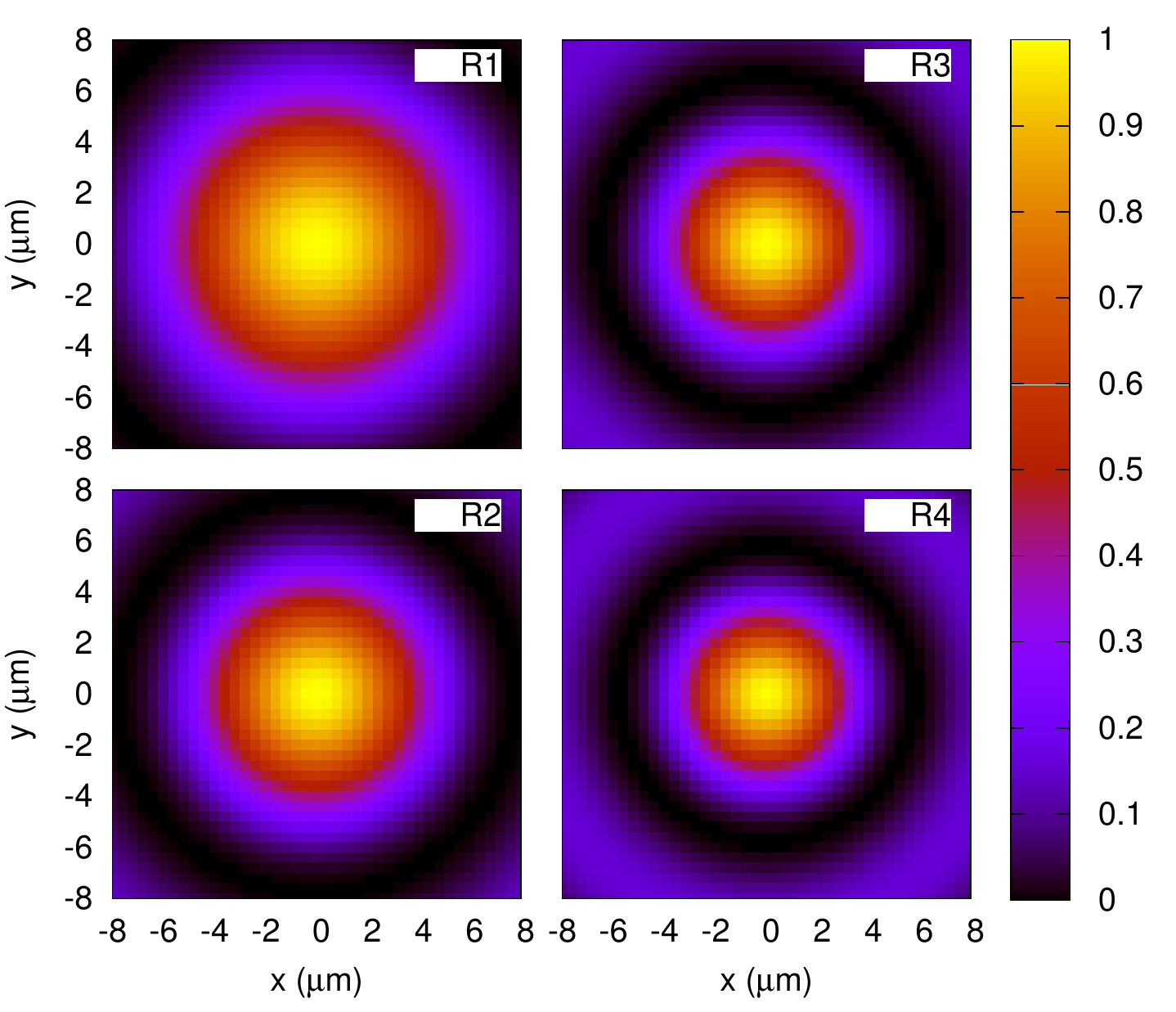}
\caption{Intensity maps in arbitrary units at the focal plane. Each map is normalized by the same $I_\text{max}$ value. Transverse profile of the focused rings is shown. Intensity peak is the same for each pulses and spot sizes are comparable.}\label{fig:maps}
\end{figure}

\begin{figure}\centering
\includegraphics[width=.9\columnwidth]{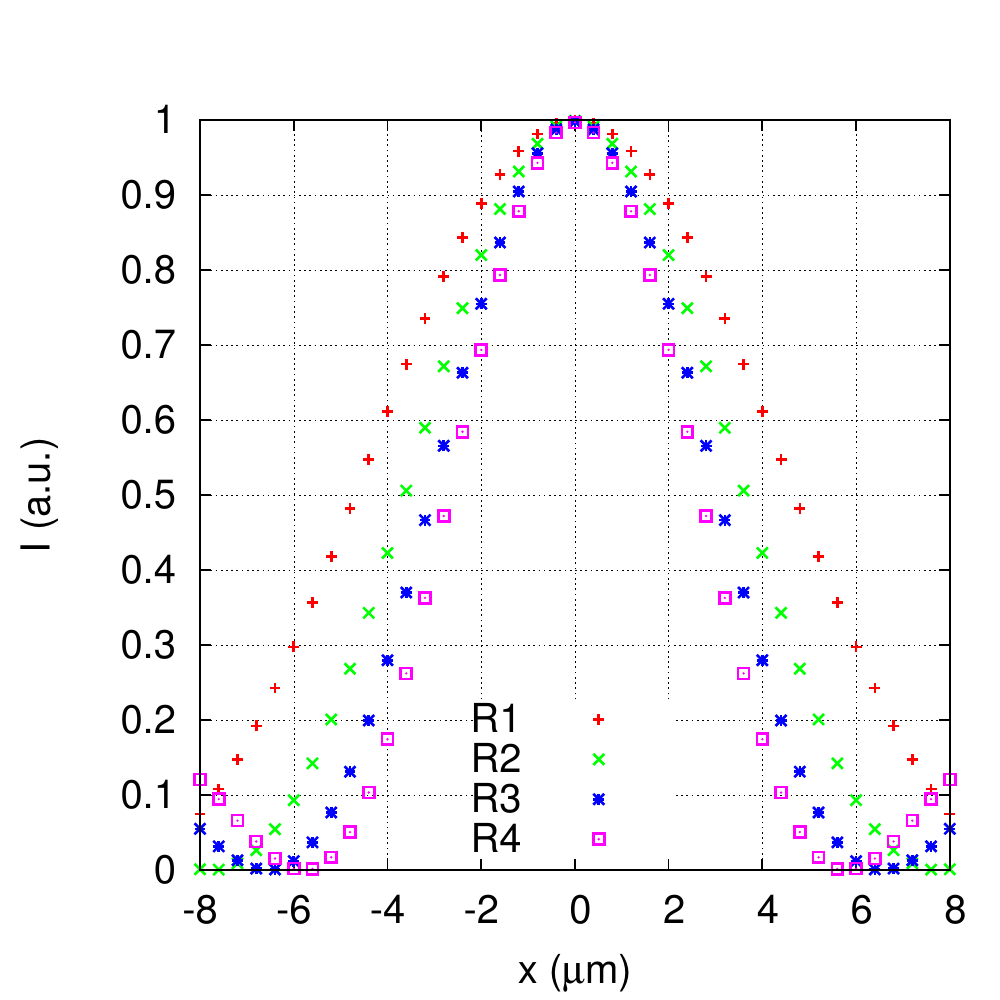}
\caption{Intensity line-out. Each plot is normalized by the same $I_\text{max}$ value. Sections with same area and incident energy result in the same peak intensity but different spot size. Diffraction effects are more relevant in the external rings.}\label{fig:space}
\end{figure}

Here we present a numerical case with the following parameters: original laser pulse $\textrm{FWHM} = \SI{40}{mm}$ (see Fig.~\ref{fig:radii}), F/5, $\theta_\textrm{OA}= \SI{25}{\degree}$, $\lambda=\SI{800}{nm}$ and $r=\SI{5}{mm}$. In Fig.~\ref{fig:maps} we plot the intensity maps at the focal plane of each train pulse. The maps show that the same-energy condition results in the same intensity peak for every pulse. In Fig.~\ref{fig:space} we plot the line-out on the $x'$ axis of the intensity for a better visualization. Spot sizes are also rather similar although diffraction effects become more important for external regions. Further studies will investigate the effects of these small differences on the excited plasma wave. We stress that these results are strictly related to the imposed conditions; working in parallel with plasma simulation can underline which one of these properties is most important or which are the best conditions to match in order to refine the mask.

\subsection{Study of the temporal properties of the train pulses}
A time dependent numerical approach including both linear and nonlinear interaction processes was undertaken in order to investigate the temporal behaviour of the generated pulse train. To this purpose we used the Mir\`o software \cite{art:miro03}. In what follows we refer to a case study aimed at a practical implementation of the ReMPI scheme.

The time distance between each pulse of the train depends solely on the group velocity. In order to resonantly excite the plasma wave the pulse separation has to match the plasma period; the thickness of each section can be set according to this requirement.

Considering a slab of fused silica, chosen for its low GVD property, with an initial electron density of \SI{e18}{cm^{-3}}, we get a difference in thickness between consecutive regions $\Delta d$ of \SI{71.52}{\micro m}. With modern manufacturing technology it is possible to produce self-standing fused silica slabs, suited for the used parameters, with a minimum thickness of \SI{500}{\micro m}; the second column of Tab.~\ref{tab:dur} shows the resulting values for the thickness of each ring.

The effects of the propagation through the mask on the time and spectral shape was simulated, using the Mir\`o code with the quasi-1D numerical scheme, for a pulse with \SI{1}{J} energy and a duration (FWHM of the intensity) of \SI{30}{fs}; different runs were carried out for each thickness $d_i$, corresponding to each mask's regions.

Since we are dealing with ultra short, intense laser pulses, both nonlinear and dispersive (up to the second order) effects were taken into account in these simulations. The results obtained for the final duration (that is, the pulse duration after passing through the mask) are reported in the third column of Tab.~\ref{tab:dur}; as it can be seen, a difference in duration between the pulses emerging from the inner ring and the outer one of approximately 10\% arises as a consequence of the propagation. Preliminary QFluid plasma simulations show that this broadening is utterly acceptable in terms of wake excitation. It is possible to mitigate this effect by adding a tiny chirp to the incoming beam; for instance, a negative frequency chirp of the order of \SI{e27}{s^{-2}} results in the durations reported in the fourth column of Tab.~\ref{tab:dur}. We notice that these results also suggest a low impact of the nonlinear effects occurring as a consequence of the propagation through the mask. Finally, we mention that the estimated $B$-integral after the mask is of order of \num{e-1}; this confirms that the thicknesses considered here can be safely tolerated in any practical situation.

\begin{table}\centering
\caption{Train pulses duration. Considering an initial pulse with a duration of \SI{30}{fs} we simulated the GVD-related time broadening, namely the FWHM of the intensity, of the train pulse emerging from each region with the corresponding thickness $d_i$. The fourth column shows the duration broadening for an initial pulse with a \SI{-1e27}{s^{-2}} frequency chirp.}\label{tab:dur}
\begin{tabular}{cccc}
\toprule
Region & $d_i$ (\si{\micro m}) & FWHM (\si{fs}) & FWHM$_c$ (\si{fs}) \\
\midrule
1 & 714 & 38.5 & 35.6\\
2 & 643 & 37.2 & 34.6\\
3 & 571 & 36.0 & 33.7\\
4 & 500 & 34.8 & 32.8\\
\bottomrule
\end{tabular}
\end{table}

\section{Conclusions}
We presented an experimental method to generate a pulse train starting from a single high-energy femtosecond pulse and using a delay mask made up by concentric rings with different thicknesses.

Diffraction effects were studied and numerical approach was developed to study the transverse profile of each train pulse allowing a fine tuning of the rings' size for the optimization of the ReMPI scheme. We showed that pulse duration broadening is not a concern for typical plasma parameters of LWFA. Further combined optical and plasma simulations are needed to improve tuning of the mask in order to enhance multi-pulse wake excitation and provide a valid experimental set up.

\section*{Acknowledgements}
This project has received funding from the European Union's Horizon 2020 research and innovation programme under grant agreement No. 653782 (EuPRAXIA project), and from the MIUR funded Italian research Network ELI-Italy. We also acknowledge funding from the Italian MIUR through the PRIN project ``Preclinical Tool for Advanced Translational Research with Ultrashort and Ultraintense X-ray Pulses'' (prot. 20154F48P9).

\bibliographystyle{elsarticle-num}

\end{document}